\documentclass[prl,preprint,aps,showpacs]{revtex4}
\usepackage{amsmath}
\usepackage{graphicx}

\def\simge{
    \mathrel{\rlap{\raise 0.511ex
        \hbox{$>$}}{\lower 0.511ex \hbox{$\sim$}}}}
\def\simle{
    \mathrel{\rlap{\raise 0.511ex
        \hbox{$<$}}{\lower 0.511ex \hbox{$\sim$}}}}
\def\slashchar#1{\setbox0=\hbox{$#1$}           
   \dimen0=\wd0                                 
   \setbox1=\hbox{/} \dimen1=\wd1               
   \ifdim\dimen0>\dimen1                        
      \rlap{\hbox to \dimen0{\hfil/\hfil}}      
      #1                                        
   \else                                        
      \rlap{\hbox to \dimen1{\hfil$#1$\hfil}}   
      /                                         
   \fi}

\begin{document}
\bibliographystyle{apsrev}



\preprint{RBRC-296}
\title{Lattice calculation of the lowest order hadronic 
contribution to the muon anomalous magnetic moment}

\author{
T.~Blum}
\affiliation{
\vspace{0.5in}
RIKEN BNL Research Center,
Brookhaven National Laboratory,
Upton, NY 11973}

\date{\today}

\begin{abstract}
We present a quenched lattice calculation of the lowest order (${\cal
O}(\alpha^2)$) hadronic contribution to the anomalous magnetic moment 
of the muon which arises from the hadronic vacuum polarization. 
A general method is presented for computing entirely in Euclidean space,
obviating the need for the usual dispersive treatment which relies on
experimental data for $e^+e^-$ annihilation to hadrons.
While the result is not yet of comparable precision
to those state-of-the-art calculations, 
systematic improvement of the quenched lattice
computation to this level is 
straightforward and well within the reach of present computers. Including
the effects of dynamical quarks is conceptually trivial, 
the computer resources required are not.
\end{abstract}


\maketitle
\newpage

The magnetic moment of the muon is defined by the $q^2\to0$ (static) limit of
the vertex function which
describes the interaction of the electrically charged muon with the photon,
\begin{eqnarray}
\Gamma_\rho(p_2,p_1)
&=&
\gamma_\rho\,F_1(q^2) - 
\frac{i}{4\,m_\mu}(\gamma_{\rho}\,\slashchar{q}-\slashchar{q}\gamma_{\rho})\,
F_2(q^2),
\label{eq:vertex}
\end{eqnarray}
where $m_\mu$ is the muon mass,
$q=p_2-p_1$ is the photon momentum, and $p_1,\, p_2$ are the incoming
and outgoing momentum of the muon. Lorentz invariance and current
conservation have been used in obtaining Eq.~\ref{eq:vertex}.
Form factors $F_1(q^2)$ and $F_2(q^2)$ contain all information about 
the muon's interaction with the electromagnetic field. In particular,
$F_1(0) = 1$ is the electric charge of the muon in units of $e$, and
$g = 2\,F_1(0) + 2\,F_2(0) = 2+ 2\,F_2(0)$
is the Land\'e g-factor, proportional to the magnetic moment. 
The anomaly, defined as half of the difference of $g$
from its tree level value, which the
Dirac equation predicts to be 2 for an elementary spin 1/2 particle, is
$a_\mu= F_2(0)$.
Thus, $F_2(0)=0$ at tree level, and corrections to $F_2(0)$, 
and therefore $a_\mu$, start 
at ${\cal O}(\alpha)$ in QED where $\alpha~=~e^2/4 \pi$ is the fine structure
constant. $F_1(0)=1$ to all orders due to charge conservation.

\begin{figure}
\includegraphics[width=\hsize]{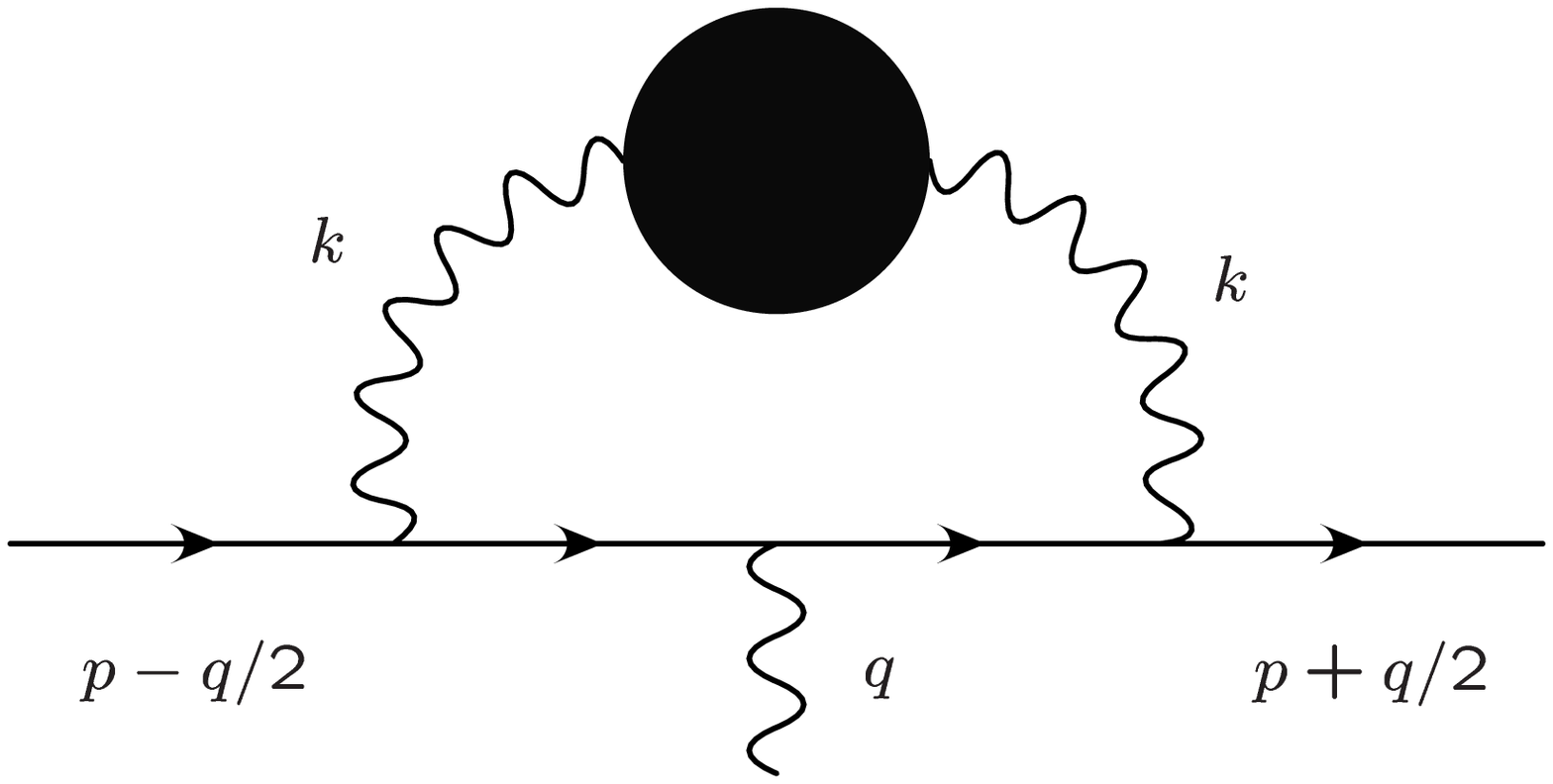}
\caption{The lowest order hadronic contribution to the muon anomalous magnetic
moment\protect\footnote{There is an additional diagram consisting of two quark
loops, each with a single insertion of the electromagnetic current, connected
by three gluons at lowest order in QCD. This diagram vanishes when summed over
three degenerate quarks $u$, $d$, and $s$ with $Q=2/3$, -1/3, and -1/3,
respectively which is the case studied here. Therefore, its contribution is
expected to be small, though this will be checked in future calculations.}.
The muon has outgoing momentum $p+ q/2$ after scattering from a photon with
momentum $q$. The loop momentum is $k$.  The blob represents the
non-perturbative hadronic vacuum polarization.}
\label{fig:diagram}
\end{figure}

The most precise measurement ever of
the muon's anomalous magnetic moment was recently carried out at Brookhaven
National Laboratory\cite{Bennett:2002jb}. In
\cite{Davier:2002dy,Hagiwara:2002ma} the authors quote 
three standard deviation discrepancies between
the Standard Model and experiment\cite{Bennett:2002jb}.
The theoretical and experimental uncertainties are roughly the
same and were added in quadrature. The dominant
theoretical uncertainty resides in hadronic loop corrections arising from  
the hadronic vacuum polarization (${\cal O}(\alpha^2)$) 
(see Figure~\ref{fig:diagram}) and hadronic
light-by-light scattering (${\cal O}(\alpha^3)$), and it
is clearly of interest to reduce these errors. 
Presently, the ${\cal O}(\alpha^2)$
hadronic contribution is calculated by using a dispersion relation and 
the experimental value of the total cross-section for $e^+\,e^-$
annihilation to hadrons to relate the 
imaginary part of the vacuum polarization to the real
part. This calculation is very precise, though a
discrepancy with a calculation that uses $\tau$ decay data may indicate
a theory error as large as five percent\cite{Davier:2002dy} and reduces the
disagreement with experiment to roughly 1.6 standard deviations. A purely
theoretical, first principles, calculation has been
lacking and is desirable, and also has several advantages over the
conventional approach. For instance, the separation of QED effects
from hadronic corrections is
automatic, as is the treatment of isospin corrections if different quark
masses are used in the simulation. 
Thus it is possible that lattice calculations may eventually help to 
settle the above mentioned discrepancy between $e^+\,e^-$
annihilation and $\tau$ decay.

The method described here is simple and direct. We begin with
Ref.~\cite{Roskies:1990ki} which describes the computation of 
multi-loop graphs in perturbation theory through the
expansion of the integrand in terms of hyperspherical polynomials.
The key is that the entire integral, including external momenta,
can be Wick-rotated into Euclidean space
and the angular integrals done so that what is left is an integral over
the magnitude of the loop momentum. If the
graph can be set up in a certain way, then after the external
momenta are analytically continued on-shell, the integral
over the loop momenta is performed without distorting the 
integration contour: $K^2=0\to\infty$, $K^2$ the squared 
Euclidean loop momentum. 
For the case at hand, this means the photon propagator can only depend
on the loop momentum. 
This is important because
if the photon propagator depends only on $K^2$, then so will
the vacuum polarization.
Thus, the renormalized hadronic vacuum polarization
function, calculated on the lattice and in Euclidean space, can be directly 
inserted into the one-loop diagram for the anomalous magnetic moment without
analytically continuing the vacuum polarization back to Minkowski
space, or requiring its value in a region not
accessible to lattice calculations, namely $K^2<0$. 
The condition on the photon propagator is easily met
in this case as can be seen from the assignment 
of momenta in Figure~\ref{fig:diagram}.

Following~\cite{Roskies:1990ki,Barbieri:1972as},
to extract $F_2(q^2)$ from Eq.~\ref{eq:vertex}, apply
a projection operator $P_\rho$
and compute ${\rm Tr}{(P_\rho\,\Gamma_\rho)}$. 
Omitting the quark loop for the moment, applying the Feynman rules, 
and taking the limit $q^2\to 0$, the
diagram in Figure~\ref{fig:diagram} gives\footnote{It is well known that
$F_2(q^2)$ is neither infrared nor ultraviolet divergent; thus we use the
unregularized photon propagator for simplicity.}
\begin{equation}
a_\mu^{(1)} = e^2\, i\, \int \frac{d^4\,k}{(2\pi)^4}
\frac{1}{((p-k)^2 + m_\mu^2 -i\,\epsilon)^2}
\frac{1}{k^2 -i\,\epsilon}
\left(\frac{16\, (p\cdot k)^2}{3\,m_\mu^2} + 
\frac{4}{3}\, k^2 + 4\, (p\cdot k)\right),
\end{equation}
where $p_{1,2} = p \mp q/2$. Analytically continuing the entire amplitude to
Euclidean space, including the external momenta,
performing the angular integrations, and then analytically 
continuing $p^2\to -m_\mu^2$ back
on-shell, we are left with
\begin{eqnarray}
a_\mu^{(1)}&=& \frac{\alpha}{\pi} 
\int_{0}^{\infty} d K^2 \frac{m_\mu^2 K^2 Z^3
(1-K^2\,Z)}{1+m_{\mu}^2 K^2 \,Z^2}=
\frac{\alpha}{\pi}\int_{0}^{\infty} d K^2 \, f(K^2),
\label{eq:schwinger}
\end{eqnarray}
where $Z= -(K^2-(K^4+4 m_\mu^2 K^2)^{1/2})/{2\,m_\mu^2\,K^2}$.
One can easily verify that this gives the leading
Schwinger contribution $a^{(1)}_\mu=\alpha/2\,\pi$. Because
the quark loop does not affect the rest of the integral, and more
importantly only depends on $K^2$, we can simply
insert its contribution into Eq.~\ref{eq:schwinger} 
to obtain the ${\cal O}(\alpha^2)$ hadronic contribution. 
\begin{equation}
a_\mu^{\rm (2)had} = \left(\frac{\alpha}{\pi}\right)^2\int_{0}^{\infty} 
d K^2 \, f(K^2)\, \hat{\Pi}(K^2),
\label{eq:had-contribution}
\end{equation}
where
$\hat{\Pi}(K^2)\equiv {4\pi^2}\sum_{i} Q^2_i\left(\Pi_i(K^2) -\Pi_i(0)\right)$.
$Q_i$ is the electric charge in units of
$e$ and $\Pi_i(K^2)$ is the vacuum polarization
for the $i^{th}$ quark flavor
which has been renormalized by subtracting its value at $K^2=0$ (from now on
we drop the subscript since we work with degenerate quarks).
In general $\Pi(K^2)$ is an analytic function of $K^2$.
In the conventional approach,
$a_{\mu}^{\rm (2)had}$ is calculated via a dispersion relation which relates
$\Re(\Pi(K^2))$ to $\Im(\Pi(K^2))$, and thus the $e^+e^-\to$ hadrons total
cross-section. In contrast, here
we deal with $\Re(\Pi(K^2))$ in the Euclidean, or space-like, region 
with real $K^2\ge0$. A simple check of this procedure, which is successful,
is to insert the one-loop QED vacuum
polarization\cite{Peskin:1995ev}, analytically continued to Euclidean space, 
into Eq.~\ref{eq:had-contribution} 
and compare the result to the known value of 
the ${\cal O}(\alpha^2)$ QED contribution to
$a_\mu$\cite{Kinoshita:1990am}. Finally, we note that the kernel 
$f(K^2)$ diverges as $K^2\to 0$,
so the integral in Eq.~\ref{eq:had-contribution} is dominated by the 
low momentum region. We now turn to the lattice calculation of $\Pi(K^2)$. 

The vacuum polarization tensor for a single quark flavor 
with unit charge is defined as
\begin{eqnarray}
\Pi^{\mu\nu}(q) &=& i\int {\rm d}^4x \,{\rm e}^{i\,q\,(x-y)}\,
\langle T\,J^\mu(x)J^\nu(y)\rangle
\end{eqnarray}
where $J^\mu(x)=\bar\psi(x)\gamma^\mu\psi(x)$ is the electromagnetic current
and $\langle\,\rangle$ signifies an average over gauge
and fermion fields.

In the lattice regularization using 
domain wall fermions\cite{Kaplan:1992bt,Shamir:1993zy}, current conservation
is given by
$\Delta^\mu J^\mu(x)=0$ where 
$\Delta^\mu$ is the backward difference operator and
\begin{eqnarray}
J^\mu(x) &=& \frac{1}{2}\sum_{s}
\bar\psi(x+\hat{\mu},s)U^\dagger(x)(1+\gamma^\mu)\psi(x,s)
-\bar\psi(x,s)U(x)(1-\gamma^\mu)\psi(x+\hat{\mu},s)
\label{eq:conserved-current}
\end{eqnarray}
is the conserved vector current\cite{Furman:1995ky}\footnote{For domain wall
fermions, we use the conventions in \cite{Blum:2000kn}.}.  The sum in
Eq.~\ref{eq:conserved-current} is over an extra fictitious 5$^{th}$ dimension
which gives rise to $4d$ Dirac fermions. These Dirac fermions are chirally
symmetric up to violations which are exponentially small in the size of this
dimension.  The Ward-Takahashi identity for the two-point function yields
\begin{eqnarray}
\label{eq:2ptward}
&&\Delta^\mu J^\mu(x)\,(J^\nu(y))^\dagger =\\\nonumber
&-&\frac{1}{2}\sum_{s}\delta(x-y)\left(
\bar\psi(y+\hat{\nu},s)U^\dagger(y)(1-\gamma^\nu)\psi(y,s)+
      \bar\psi(y,s)U(y)(1+\gamma^\nu)\psi(y+\hat{\nu},s)\right)\\\nonumber
&+&\delta(x-y-\hat{\nu})\left(
\bar\psi(y+\hat{\nu},s)U^\dagger(y)(1-\gamma^\nu)\psi(y,s)+
      \bar\psi(y,s)U(y)(1+\gamma^\nu)\psi(y+\hat{\nu},s)\right)
\end{eqnarray}
which is valid for each gauge field configuration. The contact terms
do not cancel each other because 
$J^{\mu}$ is a point-split current.
After subtracting\cite{Gockeler:2000kj}
\begin{eqnarray}
&&\delta^{\mu\nu}\sum_{s}\frac{1}{2}
\left(\right.\bar\psi(y+\hat{\nu},s)U^\dagger(y)(1-\gamma^\nu)\psi(y,s)
+\,\bar\psi(y,s)U(y)(1+\gamma^\nu)\psi(y+\hat{\nu},s)\left.\right)
\label{eq:contact-current}
\end{eqnarray}
from the two-point function to cancel the contact 
terms in Eq.~\ref{eq:2ptward}, Fourier transformation yields
the Ward-Takahashi identity
$\hat{q}^\mu\Pi^{\mu\nu}(\hat{q}) = 0$.
The polarization tensor calculated in Euclidean space, on the lattice, is 
$\Pi^{\mu\nu}(\hat{q}) =
( \hat{q}^\mu \hat{q}^\nu - \hat{q}^2 \delta^{\mu\nu})\,\Pi(\hat{q}^2)$
which follows from Euclidean and gauge invariance. The lattice
four-momentum is  
$\hat{q}=2/a\, \sin{(q^\mu/2)}$ ($q^\mu=2\pi {\rm n}_\mu/N_\mu,\,
n_\mu=0,1,\dots,N_\mu-1)$. The Ward-Takahashi identity, which
is valid before averaging over gauge fields, has been verified
to numerical precision in the present calculation (this is a good
check that the computer simulation is correct). After subtracting the
contribution of unphysical heavy fermions from a 5$^{th}$ dimension of size
$L_s$ sites\cite{Furman:1995ky},
$\hat{\Pi}(K^2)$ differs from the continuum vacuum polarization by terms that
are ${\cal O}(a^2)$.

We have calculated $\Pi(K^2)$ in the quenched approximation using the
DBW2 gauge action\cite{Takaishi:1996xj} and valence domain wall
fermions. Two values of the gauge coupling were chosen which 
correspond to inverse lattice spacing's (set from the $\rho$ meson mass), 
$a^{-1}=1.3$ and 1.96 GeV (see \cite{Aoki:2002vt}). The lattice
volumes studied were $N_s^3\times N_t=8^3\times24$ (1.3 GeV only) 
and $16^3\times32$ (1.3 and 1.96 GeV), corresponding to 
spatial volume $V=(1.2\, \rm fm)^3$, $(2.4\, \rm fm)^3$, and 
$(1.6\, \rm fm)^3$, respectively. For the domain wall fermions we used
$L_s=8$, domain wall height $M_5=1.8$,  
and a single $4d$ quark mass $m_f=0.04$, or roughly 90 and 120
MeV in the $\overline{MS}$ scheme at $\mu=1.3$ and 1.96 GeV,
respectively\cite{Dawson:2002nr}. 

\begin{figure}
\includegraphics[width=\hsize]{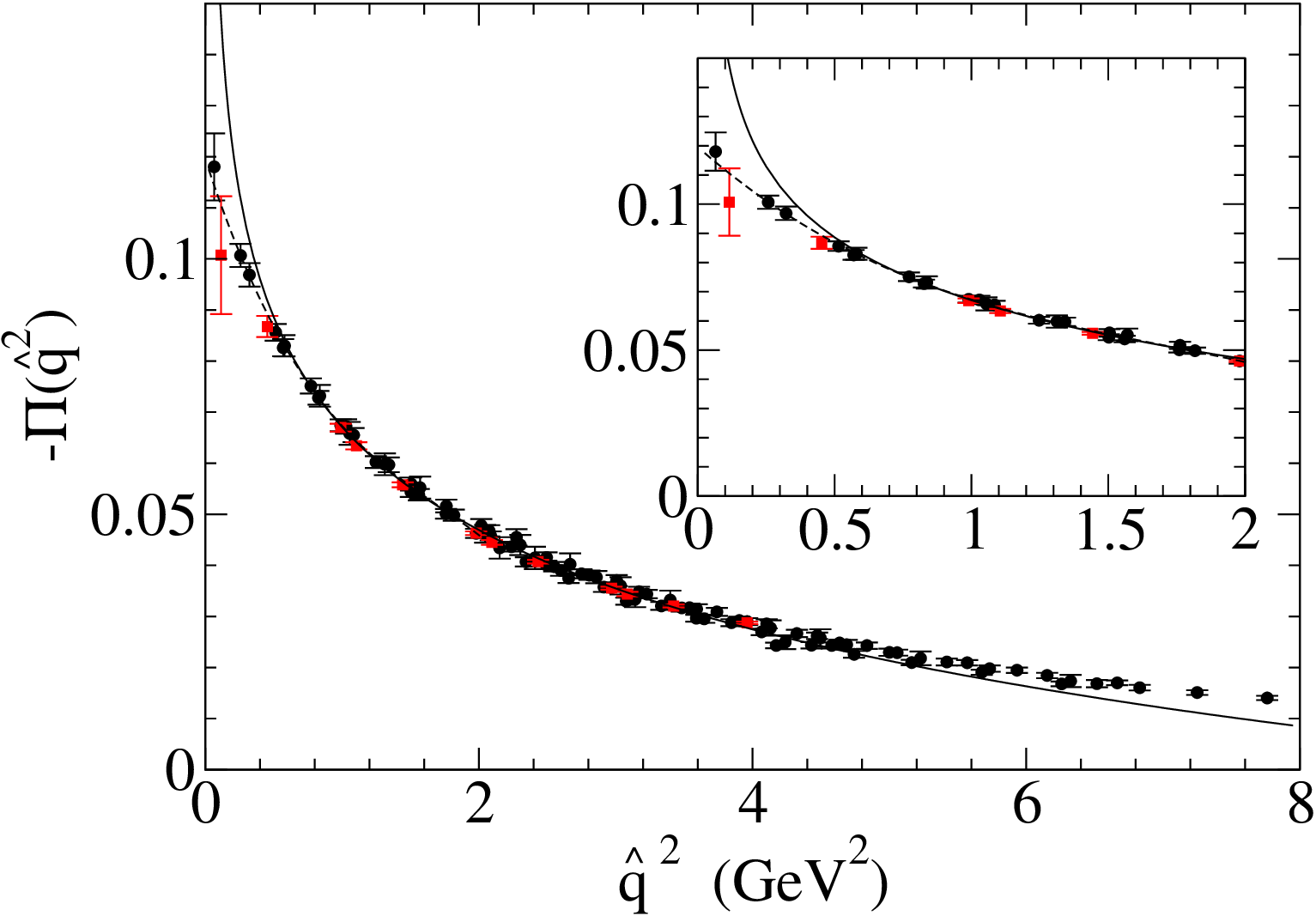}
\caption{The unsubtracted vacuum polarization for the $a^{-1} =1.3$
GeV lattice. The two volumes
described in the text are shown, $8^3$ (squares) and $16^3$ (circles).
The solid line denotes a three-loop perturbation theory 
calculation\cite{Chetyrkin:1996cf},
which has been shifted by a constant for comparison, and the dashed line the
fit of the larger volume data points.
}
\label{fig:vacpol-1.3}
\end{figure}

Shown in Figure~\ref{fig:vacpol-1.3} is the vacuum polarization for the large
volume at $a^{-1}=1.3$ GeV, the most physically interesting one since we are
mainly interested in the small $\hat{q}$ region. The agreement with
perturbation theory\cite{Chetyrkin:1996cf} is very good, even down to
$\hat{q}^2\sim 0.5$ GeV$^2$, until finally non-perturbative effects begin to
be important. For very large values of $\hat{q}$ lattice artifacts dominate,
and the lattice and continuum results disagree.  The perturbative result is
evaluated in the $\overline{MS}$ scheme at $\mu=1/a=1.3$ GeV with a quark mass
$m_q\approx 90$ MeV\cite{Dawson:2002nr} and has been shifted by a
constant\footnote{ The lattice and perturbative results are computed in
different regularization schemes, and the logarithmic divergence has not been
subtracted from the former.}  in Figure~\ref{fig:vacpol-1.3} for comparison.
We also show the small volume results which indicate finite volume effects are
negligible until $\hat{q}^2\simle 0.5$ GeV$^2$. The results from the $1.96$
GeV lattice are quite similar, indicating that non-zero-lattice-spacing errors
are small.

Since we are interested in the low $\hat{q}^2$ region,
to use Eq.~\ref{eq:had-contribution}, we fit the lattice data to
a simple polynomial in $\hat{q}^2$, which allows for a smooth 
interpolation of the data.
This amounts to a Taylor expansion about $\hat{q}^2=0$.  Lorentz covariance
(really, hyper-cubic symmetry) requires $\Pi(\hat{q}^2)$ depend only on
$\hat{q}^2$, and if the quark mass is non-zero, it must be regular as
$\hat{q}^2\to0$.  A more sophisticated anastz is clearly desirable, especially
one that includes finite volume effects. However, the construction of such an
ansatz, for example in chiral perturbation theory, is outside the scope of
this work.
The value of $\Pi(\hat{q}^2)$ between $\hat{q}^2=0$ and 
the smallest value calculated on the lattice
is extrapolated from the fit, so it is important to have
values of $\hat{q}^2$ that are close to zero. The smallest value
in this study, which is on the large volume, is 
$\hat{q}^2_{\rm min} \approx 0.065$.
Specifically, we use a four parameter fit function,
$\Pi(x) = a_0 + a_1\, x + a_2\, x^2 + a_3\, x^3$
where $x=\hat{q}^2$, and fit the data in the range $x\le 2$ GeV$^2$. The fit is
uncorrelated but performed under a jackknife procedure. The final
results are insensitive to the number of parameters used and the fit
range, so long as the chosen combination accurately reproduces
the data. That is, the larger the range, the more parameters needed 
to accurately represent the low momentum region 
(see Figure~\ref{fig:vacpol-1.3}).

The fit function is next plugged into Eq.~\ref{eq:had-contribution} and
integrated numerically, using MATHEMATICA, up to some cut, $\hat{q}^2_{\rm
cut}$.  The perturbative value of $\Pi(K^2)$ is then used in the integral from
$\hat{q}^2_{\rm cut}$ to $\infty$, after it has been shifted by a constant to
match onto the subtracted lattice result.  The final result is quite
insensitive to the value of the cut since $f(K^2)$ in
Eq.~\ref{eq:had-contribution} is sharply peaked at zero, and with the
statistical error on the present data such as it is, the perturbative
contribution can be ignored entirely since it adds, depending on the cut,
$(1-10)\times 10^{-10}$ to $a_\mu^{\rm (2)had}$.  However, as the lattice
results become more accurate, these contributions will have to be carefully
included. To be specific, results are given for $\hat{q}^2_{\rm
cut}=1.5$~GeV$^2$, though a much smaller value yields essentially the same
answer.

Using the above procedure and including degenerate $u$, $d$, and $s$ quarks we
find $a_\mu^{\rm (2)had}= 460(78)\times10^{-10}$ on the large volume (the
error is statistical). This is roughly 2/3 of the value computed using the
dispersive approach\cite{Davier:2002dy}, and given the approximations in this
first calculation: quenching, finite volume, and unphysically large quark
masses, quite encouraging. 
We note with respect to the quenching systematic error, the above result is
quite reasonable: 72\% of the dispersive result comes directly from the $\rho$
resonance\cite{Davier:2002dy}, essentially all that is included in the
quenched case.
The result depends heavily on the low $\hat{q}^2$
region, so the final statistical error is still rather large since only a
small number of configurations were used (27) to compute averages over the
gauge-field. The smaller volume result ($a^{-1}=1.3$ GeV, 138 configurations)
is roughly $a_\mu^{\rm (2)had}= 318(69)\times10^{-10}$, indicating large
finite volume effects. The 1.96 GeV lattice, which corresponds to a somewhat
larger volume, gives $a_\mu^{\rm (2)had}= 378(96)\times10^{-10}$ (18
configurations), in between the large and small 1.3 GeV lattices.

Since the hadronic loop is given by a sum over all possible values of the
momentum of the quarks, one should worry that lattice artifacts due to finite
$L_s$ may be noticable. Such effects were observed in domain wall fermion
calculations of the chiral condensate and weak matrix
elements\cite{Blum:2000kn,Blum:2001xb} and are well understood.  In the latter
case, the physical matrix element was obtained by taking a slope with respect
to quark mass which easily removed the lattice artifact. A similar mechanism
is at work here.  $\Pi(\hat{q}^2)$ depends on $L_s$ through a constant,
momentum independent shift which vanishes as $L_s\to\infty$, but the
renormalized $\hat{\Pi}(\hat{q}^2)$ is independent of $L_s$ to a very high
degree. This was checked by calculating $a_\mu^{\rm (2)had}$ with $L_s=4$ on
the small volume, 1.3 GeV lattice. While the values of $\Pi(0)$ differ by
roughly 30\%, $a_\mu^{\rm (2)had}= 317(64)\times10^{-10}$ calculated with
$L_s=4$ is in excellent agreement with the $L_s=8$ value. Being purely an
issue of the physics near the ultraviolet cut-off where the heavy 5$d$
fermions play a role (apart from the small additive quark mass, $m_{\rm
res}$\cite{Blum:2000kn}, which affects the low energy physics), this could
have been anticipated from Figure~\ref{fig:vacpol-1.3}.  Since the agreement
with perturbation theory is so good, visible effects would only show up as a
constant.

We note that the vacuum polarization can also be fit to a form given by the
operator product expansion\cite{Gockeler:2000kj}, though being a short
distance expansion, it is valid for $\hat{q}^2\gg 1$.  In particular, in this
(quenched) study we find no evidence for power corrections beyond the operator
product expansion, which is in agreement with \cite{Gockeler:2000kj}.

We are optimistic about the prospects for improving this calculation.  Further
quenched calculations on larger volumes will pin down finite volume effects,
reduce the statistical error (there is room for significant improvement), and
probe the chiral limit for the light quarks.  The disconnected diagram that
does not vanish in the non-degenerate case will also be computed.  This should
reduce the error on the quenched calculation below five percent.  Recent
simulations with dynamical domain wall fermions\cite{Izubuchi:2002pt} may be
valuable in estimating the quenching error. 
Perhaps even more interesting in the near term, recent 2+1 flavor dynamical
fermion calculations using improved Kogut-Susskind fermions\footnote{Private
communication with Doug Toussaint of the MILC
collaboration.}\cite{Bernard:2002ep} offer a promising avenue to address all
the main systematic errors in this first calculation.
The procedure set down here goes through
in just the same way, but with dynamical gauge fields replacing the quenched
ones. It is the generation of gauge fields with the dynamical fermion effects,
especially for light quarks, that is so costly. Though we are still a ways
from competing with the precision of the dispersive method, lattice
calculations should eventually rival that precision. We have already mentioned
that the lattice method avoids the problems of disentangling QED effects from
the hadronic corrections, and isopin breaking can be handled simply by using
different quark masses for the $u$ and $d$ quarks. In any case, a completely
theoretical, first principles calculation, though challenging, is well worth
the effort.

\bibliography{paper}


\section*{Acknowledgments}

I am grateful to W. Marciano for many useful discussions, and in
particular for pointing out the work in Ref.~\cite{Roskies:1990ki}. 
I also thank M. Creutz, T. Izubuchi, and A. Soni for helpful discussions.
I thank RIKEN, Brookhaven National Laboratory and the U.S.  Department
of Energy for providing the facilities essential for the completion of
this work. All computations were carried out on the QCDSP supercomputers at 
the RIKEN BNL Research Center and Columbia University.



%
%
\newpage

\newpage

\end{document}